\documentstyle[preprint,aps,prb,epsfig]{revtex}
\begin{document}
\draft
\tightenlines

\title{Magnetoplasmon excitations in arrays of circular and noncircular
quantum dots}

\author{B.P.~van Zyl and E.~Zaremba}
\address{Department of Physics, Queen's University,\\ Kingston, 
Ontario, Canada K7L 3N6}
\author{D.~A.~W.~Hutchinson}
\address{Clarendon Laboratory, Parks Road, \\University of Oxford, \\ Oxford,
England}
\date{\today}

\maketitle
\begin{abstract}

We have investigated the magnetoplasmon excitations in arrays of 
circular and noncircular quantum dots within the 
Thomas-Fermi-Dirac-von Weizs\"acker approximation.
Deviations from the ideal collective excitations of isolated 
parabolically confined electrons arise from local perturbations of the
confining potential as well as interdot Coulomb interactions. The latter
are unimportant unless the interdot separations are of the order of the
size of the dots. Local perturbations such as radial anharmonicity and
noncircular symmetry lead to clear signatures of the violation of the
generalized Kohn theorem. In particular, the reduction of the local
symmetry from $SO(2)$ to $C_4$ results in a resonant coupling of
different modes and an observable anticrossing behaviour in the power
absorption spectrum. Our results are in good agreement with recent
far-infrared (FIR) transmission experiments.

\end{abstract}

\newpage

\section{INTRODUCTION}

Advances in nanofabrication technology have opened up the
possibility of studying electronic systems under a wide range of
confinement conditions. The systems of interest are usually
fabricated by starting with a two-dimensional electron gas and applying
various lateral modulation 
techniques.\cite{mackens1984,heitmann1993,davies1997} In this way,
systems such as wires, dots, rings and antidot arrays have been
realized. A powerful method for studying these systems is FIR 
absorption spectroscopy.  The observation of collective plasmon
excitations and their dispersion in an 
applied magnetic field can yield a great
deal of information about the effects of electron-electron interactions
and the influence of different forms of geometrical confinement.

In this paper we are primarily interested in the magnetoplasmon 
excitations in quantum dot arrays. In particular, we focus
on the experimental results of Demel 
{\it et al.}~\cite{demel1990,demel1990b} which reveal
interesting features in the absorption spectrum as a function of
magnetic field. It is by now well understood that the observed 
deviations from the ideal spectrum expected for parabolic 
confinement~\cite{nazin1989,maksym1990,shikin1991}
are mainly due to anharmonic perturbations of the 
confining potential.~\cite{gudmundsson1991,pfannkuche1991,nazin1992,ye-zaremba1994} 
For parabolic confinement, the only allowed
dipole excitations correspond to center of mass motion and have
the frequencies $\omega_\pm = \sqrt{\omega_0^2 + (\omega_c/2)^2}
\pm \omega_c/2$, where $\omega_0$ is the harmonic confining
frequency and $\omega_c$ is the cyclotron frequency.
In a previous paper \cite{ye-zaremba1994} it was shown that an
axially symmetric $r^4$ perturbation leads to a coupling of the
center-of-mass mode to other
dipole excitations of the harmonic potential, and gives rise to
a satellite peak which tracks along the $\omega_+$ excitation. The
calculated absorption spectrum was found to be in good agreement
with experiment with regard to both the position and strength of
the satellite peak.

One feature which this earlier calculation~\cite{ye-zaremba1994}
did not account for was an
anticrossing behaviour observed at lower magnetic fields. It
was originally conjectured~\cite{demel1990,demel1990b}
that this feature was due to the
nonlocal dynamic response associated with the finite
compressibility of an electron gas and responsible for the
wavevector dispersion of plasmons in a uniform system.
An explanation on the basis of a noncircular confining potential
was dismissed on the grounds that the observed anticrossing did
not conform with its expected dependence on $N$, the number of
electrons in the dot. Subsequent work, however, shifted towards
the noncircular confining potential as being the most likely
explanation of the observed behaviour. Gudmundsson and Gerhardts
\cite{gudmundsson1991} 
performed random-phase approximation (RPA) calculations for a
parabolic dot with $10 \le N \le 30$ and argued, on the basis of
the coupling between the center of mass and relative degrees of
freedom induced by a $x^2y^2$ perturbation, that an anticrossing
of the dipole mode with modes of higher multipolarity is to be
expected. Pfannkuche and Gerhardts \cite{pfannkuche1991} arrived at 
similar conclusions based on an exact diagonalization of the dot
Hamiltonian with quartic perturbations, but only for $N=2$. A
classical electrostatic model was used by Nazin {\it et al.}
\cite{nazin1992}
to study the magnetoplasmon modes in a parabolic dot with a $r^4
\cos(4\theta)$ perturbation. This approach, which treats the
electrons as a charged classical fluid, can be formulated in terms of
an approximate energy functional which includes only the effects 
of the confining potential and the electrostatic repulsion between 
the electrons. The model is useful for large $N$ and was shown
to provide a good description of the dynamics in anharmonic
dots.~\cite{ye-zaremba1994}
However, the theory proved to be ill-defined for
noncircular dots as a result of the singular behaviour of the
equilibrium density at the edge of the dot,~\cite{nazin1992} and a 
rigorous analysis of the dynamics could not be provided.

Our purpose here is to re-examine circular and noncircular quantum dots within
the Thomas-Fermi-Dirac-von Weizs\"acker (TFDW) hydrodynamic theory
developed in the context of parabolically confined electron slabs.
\cite{zaremba-tso1994}
The theory was later applied to electron rings \cite{zaremba1996} 
and was shown to give a good account of the edge magnetoplasmon modes
observed in this system. The theory improves on the classical
electrostatic model in that the quantum kinetic energy is
included at the level of the TFDW
approximation. In addition, exchange-correlation effects can also be
included.  Unlike other hydrodynamic-based models,
\cite{fetter1974,eliasson1986a,eliasson1986b,lai1986,cataudella1988} 
the TFDW hydrodynamic theory incorporates the calculated equilibrium 
properties of the electronic density and thereby provides
a more physical description of the collective modes.

The approach that we use is best suited to treating the FIR response of
dots with a large number of electrons where the response is expected to
be dominated by collective-like excitations. For few electrons, the
underlying electronic structure can play an important role. In this
situation, methods based on the random phase approximation are more
appropriate, and calculations for noncircular dots in this regime have
recently appeared.\cite{ullrich1999,mag1999} However, these 
calculations cannot be extended readily to dots with a large number of 
electrons which is the regime of interest in this paper.

Our paper is organized as follows. In Sec. \ref{tfdw_formalism} we 
present an overview of the TFDW formalism and develop the constitutive
equations for the theory in the context of a bi-directionally
modulated electronic system.  
In Sec. \ref{quantumdots}, 
we apply our theory to dot arrays and examine the effects of
confinement symmetry and interdot coupling
on the magnetoplasmon modes of the system.
Finally, in Sec. \ref{conclusions},
we present our concluding remarks.
A brief account of this work will appear as a conference 
proceeding.\cite{vanzyl99a}

\section{THE TFDW FORMALISM}
\label{tfdw_formalism}

The equilibrium and hydrodynamic properties of a periodically modulated
2DEG within the TFDW approximation  have
already been discussed at some length in our previous
work on uniaxially modulated electronic systems.~\cite{vanzyl99}
In this section, we provide an overview of
the TFDW formalism and extend our earlier work to the more general
case of bi-directional modulation.   Since
these calculations parallel those in Ref.~[\onlinecite{vanzyl99}], 
we only present the essential ingredients of the model here and refer 
to our earlier papers for details.~\cite{zaremba-tso1994,vanzyl99} 

\subsection{Equilibrium Properties}
\label{equilibrium}

As in all density functional theory (DFT) schemes, \cite{kohn1983}
the equilibrium properties of the system
are obtained by finding the variational minimum of an energy functional.
In atomic units ($e^2/\epsilon = m^{\star} = \hbar = 1$), the TFDW 
functional is given by  
\begin{eqnarray}
E[n] &=& \int d^2{\bf r} \left[ \frac{\pi}{2} n^2 + \frac{\lambda_w}{8} 
\frac{\mid \nabla n(\bf{r}) \mid^{2}}{n(\bf{r})} - 
\frac{4}{3}\sqrt{\frac{2}{\pi}} n^{3/2} \right] \nonumber \\ &+&
\frac{1}{2} \int d^2{\bf r} \int d^2{\bf r}'
\frac{n({\bf r}) n({\bf r}')}{\mid {\bf r} - {\bf r}' \mid} +
\int d^2{\bf r}\, v_{\rm ext}({\bf r}) n({\bf r})\,.
\label{functional}
\end{eqnarray}
The first term in (\ref{functional}) is the Thomas-Fermi kinetic
energy, the second term is a von Weizs\"acker-like correction to the
kinetic energy,\cite{footnote1} and the third term is the
Dirac local exchange energy.   For simplicity, we
neglect any correlation contribution. Following our earlier
work,~\cite{vanzyl99} we choose a value of $\lambda_w = 0.25$ for 
the von Weizs\"acker coefficient. Our results are not strongly dependent
on this value since the systems studied are primarily in the 
Thomas-Fermi regime.

A variation of Eq.~(\ref{functional}) with respect to the density 
leads to the Euler-Lagrange
equation
\begin{equation}
\frac{\delta E[n]}{\delta n({\bf r})} - \mu = 0~,
\label{variation}
\end{equation}
where the Lagrange multiplier $\mu$ (chemical potential)
serves to fix the total number of electrons, $N$. 
Introducing the so-called von Weisz\"acker wave function, 
$\psi ({\bf r}) \equiv \sqrt{n({\bf r})}$, into Eq.~(\ref{variation})
we obtain

\begin{equation}
- \frac{\lambda_{w}}{2} \nabla^{2} \psi ({\bf r}) +
v_{\rm eff} ({\bf r}) \psi ({\bf r}) = \mu \psi ({\bf r})~,
\label{SE}
\end{equation}
where the effective potential is given by

\begin{equation}
v_{\rm eff}({\bf r}) = \pi \psi^{2}({\bf r}) - \sqrt{\frac{8}{\pi}}
\psi ({\bf r}) + \phi ({\bf r})+v_{\rm ext}({\bf r})~.
\label{veff}
\end{equation}
Here, $\phi({\bf r}) = \int d{\bf r}' n({\bf r}')/|{\bf r} -
{\bf r}'|$
is the electrostatic potential arising from the electronic
density $n({\bf r})$, and $v_{\rm ext} ({\bf r})$ is the externally
imposed potential. The latter is assumed to be periodic with
periodicities $a_x$ and $a_y$ in the $x$ and $y$ directions,
respectively, and to have inversion symmetry about the center of the
unit cell.  

Eqs.~(\ref{SE}) and (\ref{veff}) are reminiscent of the usual Kohn-Sham
equations except that only a single electronic orbital need be
calculated. It should also be noted that the effective potential $v_{\rm
eff}$ includes a term coming from the Thomas-Fermi kinetic
energy which does not appear in the usual Kohn-Sham potential.
The required solution to Eq.~(\ref{SE}) is the ground state von 
Weisz\"acker wave function, $\psi_0$, which is obtained by imposing the
boundary condition

\begin{equation}
\hat {\bf n} \cdot \nabla \psi_0 ={\bf 0}~,
\end{equation}
at the edges of the unit cell. Here, $\hat {\bf n}$ is a unit normal
vector. To complete the specification of $\psi_0$ we must also impose
the normalization
\begin{equation}
\int d^2{\bf r}~n_0({\bf r}) = N~,
\end{equation}
which fixes the chemical potential $\mu$.

The self-consistent solution of Eqs.~(\ref{SE}) and (\ref{veff}) can be
obtained by direct iteration,~\cite{vanzyl99} but care must be taken 
to avoid numerical instabilities associated with charge 
fluctuations. Here we adopt an alternative method to avoid this 
difficulty. We use the method of imaginary time evolution defined by the
equation
\begin{equation}
\dot{\psi} =  - (H-\mu)  \psi ~,
\label{steep}
\end{equation}
where $H$ is the TFDW Hamiltonian given by

\begin{equation}
H = -\frac{\lambda_{w}}{2} \nabla^2 + v_{\rm eff}~.
\end{equation}
Using a first-order 
approximation for the time derivative on the left hand side of
Eq.~(\ref{steep}), we have
\begin{equation}
\psi (t+\delta t) = \psi (t) + (\mu -H) \psi(t) \delta t~,
\label{timestep}
\end{equation}
where $\mu$ is given by
\begin{equation}
\mu \simeq \frac{\int d{\bf r}\, \psi^* H \psi}
{\int d{\bf r}\,\psi^*\psi}~.
\end{equation}

Once an initial guess for $\psi(t=0)$ has been made,
Eq.~(\ref{timestep}) provides an updated $\psi (t)$ which is used
in the next iteration to
determine an updated $H$ and $\mu$, and so on until convergence.
This scheme preserves the normalization of the wave function to
$O(\delta t^2)$, but the wave function is nonetheless normalized 
to unity at each iteration.
A combination of real-space and Fourier-space representations is used to
evaluate the effect of $H$ on $\psi$.
We have found this method of generating $\psi_0$ to be very stable
against the charge fluctuation induced Coulomb instability. 

\subsection{Hydrodynamic Equations}
\label{dynamics}

A theory for the collective excitations of a modulated 2DEG is 
constructed by treating the electronic system as an inviscid, charged 
``classical'' fluid. In particular, we adopt the hydrodynamic
equations~\cite{zaremba-tso1994,vanzyl99}
\begin{equation}
\left( \frac{\partial n}{\partial t} \right ) + 
\nabla \cdot (n{\bf v}) = 0~,
\label{continuity_2}
\end{equation}
and
\begin{equation}
n \left [
\frac{\partial {\bf v}}{\partial t} + {\bf v}\cdot \nabla {\bf v}
\right ] =
n {\bf F}^{\rm tot},
\label{momentum_2}
\end{equation}
where the total force acting on a fluid element,
${\bf F}^{\rm tot} = {\bf F}^{\rm int} + {\bf F}^{\rm ext}$, consists of
the internal force

\begin{equation}
{\bf F}^{\rm int} ({\bf r},t) = -\nabla \left [
v_{\rm eff}  ({\bf r},t) - \frac{\lambda_{w}}{2} 
\frac{\nabla^2 \psi  ({\bf r},t)}{\psi  ({\bf r},t)} \right ]~,
\end{equation}
and the force due to external electromagnetic fields ($c = 1$)
\begin{equation}
{\bf F}^{\rm ext} = -\left ( {\bf E}^{\rm ext} + {\bf v} \times
{\bf B}^{\rm ext} \right )\,.
\end{equation}

We wish to show here that these hydrodynamic equations can be obtained
within a Lagrangian formulation. We suppose that the external
electromagnetic field is defined in terms of a scalar potential 
$\phi^{\rm ext}$ and a vector potential ${\bf A}^{\rm ext}$ by
\begin{equation}
{\bf E^{\rm ext}} = -\nabla \phi^{\rm ext} -\frac{\partial
{\bf A}^{\rm ext}}{\partial t},~~~~~~~ {\bf B}^{\rm ext} = \nabla \times
{\bf A}^{\rm ext}~,
\end{equation}
and that the velocity field ${\bf v}({\bf r},t)$ is defined by

\begin{equation}
{\bf v}({\bf r},t) = \nabla g_{1}({\bf r},t) + g_{2}({\bf r},t)
\nabla g_{3}({\bf r},t) + {\bf A}^{\rm ext}~.
\label{velocity}
\end{equation}
Here, the $g_{i}$'s are three independent scalar functions; $g_1$ is
associated with the irrotational part of the fluid flow and $g_2$ and
$g_3$ are introduced to represent the solenoidal part. The dependence on
the vector potential is displayed as a separate term.

The Lagrangian density that we propose is given by

\begin{equation}
{\cal L} = n \left (
\frac{{\bf v}^{2}}{2} + \frac{\partial g_{1}}{\partial t} +
g_{2} \frac{\partial g_{3}}{\partial t} \right ) - 
n \phi^{\rm ext} + \varepsilon [n] 
\label{lagrangian}
\end{equation}
where $\varepsilon [n]$ is the TFDW energy density.
Application of the principle of least action

\begin{equation}
\delta \int dt \int d{\bf r}~{\cal L}  = 0~,
\label{least_action}
\end{equation}
for each of the variables $\{ g_1\,,g_2\,,g_3\,,n\}$ generates four 
equations. It is straightforward to show that independent variations 
of the $g_i$ yield the following equations:

\begin{equation}
\frac{\partial n}{\partial t} + \nabla \cdot (n{\bf v}) = 0~,
\label{continuity}
\end{equation}
\begin{equation}
\frac{\partial g_2}{\partial t} + {\bf v} \cdot \nabla g_2 = 0~,
\label{g2}
\end{equation}
\begin{equation}
\frac{\partial g_3}{\partial t} + {\bf v} \cdot \nabla g_3 = 0~.
\label{g3}
\end{equation}
Eq.~(\ref{continuity}) is just the expected continuity equation.

Now, a variation of Eq.~(\ref{least_action}) with respect to $n$ yields

\begin{equation}
\left(
\frac{{\bf v}^2}{2} + \frac{\partial g_1}{\partial t} +
g_2\frac{\partial g_3}{\partial t} \right ) - \phi^{\rm ext} +
\left( v_{\rm eff} - \frac{\lambda_{w}}{2}\frac{ \nabla^2\psi}{\psi}
\right) = 0.
\label{varn}
\end{equation}
Taking the gradient of Eq.~(\ref{varn}), along with the time 
differentiated form of Eq.~(\ref{velocity}), we obtain

\begin{equation}
\frac{\partial {\bf v}}{\partial t} + ({\bf v}\cdot\nabla){\bf v}
+ {\bf E}^{\rm ext} + {\bf v}\times(\nabla\times{\bf v})-
\frac{\partial g_2}{\partial t}\nabla g_3 + 
\frac{\partial g_3}{\partial t}\nabla g_2 +
\nabla \left(v_{\rm eff} - \frac{\lambda_{w}}{2} 
\frac{ \nabla^2\psi}{\psi} \right) = 0~.
\label{momentum_1}
\end{equation}
Using Eq.~(\ref{velocity}) to evaluate the triple cross product, we have

\begin{equation}
{\bf v}\times(\nabla \times{\bf v}) = ({\bf v}\cdot \nabla g_3)\nabla 
g_2 -({\bf v}\cdot \nabla g_2)\nabla g_3 + {\bf v}\times {\bf B}^{\rm
ext}~,
\label{triple_cross}
\end{equation}
and substituting this result together with
Eqs.~(\ref{g2}) and (\ref{g3}) into Eq.~(\ref{momentum_1}), we 
finally obtain 

\begin{equation}
\frac{\partial {\bf v}}{\partial t} + {\bf v}\cdot \nabla {\bf v}
= - \nabla \left( v_{\rm eff} - \frac{\lambda_{w}}{2}
\frac{ \nabla^2\psi}{\psi} \right) - \left( {\bf E}^{\rm ext} + 
{\bf v} \times {\bf B}^{\rm ext} \right )~.
\label{momentum}
\end{equation}
Eqs.~(\ref{continuity}) and (\ref{momentum}) are equivalent to the two
equations (\ref{continuity_2}) and (\ref{momentum_2}) written down at
the beginning of this section. These are the fundamental hydrodynamic
equations of the TFDW theory.

Linearizing our hydrodynamic equations about small deviations from
equilibrium, {\em viz.,} 
$n ({\bf r},t) \rightarrow n_{0} + \delta n ({\bf r},t)$, and retaining
only first-order quantities, Eqs. (\ref{continuity_2}) and (\ref{momentum_2})
yield

\begin{equation}
\frac{\partial \delta n}{\partial t} + \nabla \cdot (n_{0} {\bf v}) = 0~,
\label{lincont}
\end{equation}
and
\begin{equation}
\frac{\partial {\bf v}}{\partial t} =
\delta {\bf F}^{\rm int} - {\bf E}^{\rm ext} - {\bf v} \times 
{\bf B}^{\rm ext}~,
\label{linmom}
\end{equation}
where the fluctuating force is given by

\begin{equation}
\delta {\bf F}^{\rm int}({\bf r},t) = - \nabla \left[ \delta 
v_{\rm eff}({\bf r},t) -
\frac{\lambda_{w}}{2 \psi_{0}} \nabla^{2} \delta \psi ({\bf r},t)+
\frac{\lambda_{w}}{2} \frac{\nabla^{2} \psi_{0}}{\psi_{0}^{2}}
\delta \psi ({\bf r},t) \right] \equiv -\nabla f~,
\label{fluctforce}
\end{equation}
with

\begin{equation}
\delta v_{\rm eff} ({\bf r},t) = 2{\pi} \psi_{0}\delta \psi 
({\bf r},t) - \sqrt{\frac{8}{\pi}}\delta \psi ({\bf r},t) + 
\delta \phi ({\bf r},t)~.
\label{flucveff}
\end{equation}
The quantity $f$ plays the role of an effective potential for the
internal force fluctuation. The magnetic field ${\bf B}^{\rm ext}$ will
be taken to be a uniform field in the $z$-direction.

For the moment, we shall consider the normal modes of the system which
are determined in the absence of the driving field ${\bf E}^{\rm ext}$.
Combining Eqs.~(\ref{lincont}) and 
(\ref{linmom}) leads to the following equation for the fluctuating 
part of the von Weizs\"acker wave function ($\delta \psi = \delta
n/2\psi_0$)

\begin{equation}
\omega (\omega^2 - \omega_{c}^{2} ) \delta \psi =
-\frac{1}{2} \omega \nabla^2 (\psi_0 f) +
\frac{1}{2} \omega \left ( \frac{\nabla^2 \psi_0}{\psi_0} \right )
(\psi_0 f) + i \omega_c (\nabla \psi_0 \times \nabla f) \cdot
\hat{\bf z}~.
\label{master}
\end{equation}
Here, $\omega_c = eB^{\rm ext}/m^\star c$, is the cyclotron frequency
expressed in natural units.
Making explicit use of the periodicity of the system in both
spatial directions, we can write the fluctuating part of the
von Weisz\"acker wave function in the form

\begin{equation}
\delta \psi ({\bf r},t) = e^{i ({\bf q} \cdot {\bf r} - \omega t)} 
%\sum_{{\bf G}} c_{{\bf G}} \varphi_{{\bf G}}({\bf r})~,
\sum_{{\bf G}} c_{{\bf G}} e^{i {\bf G} \cdot {\bf r}}~, 
\label{fourier_psi}
\end{equation}
where $c_{{\bf G}}$ is a Fourier expansion coefficient, ${\bf q}$ is a
2D Bloch wavevector restricted to the first Brillouin zone, and
${\bf G} = (2\pi m/a_x ,2\pi n/a_y )$, with $m,n = 0, \pm 1, \pm 2$, 
..., are reciprocal lattice vectors.  The plane wave basis functions
%$\varphi_{{\bf G}} = {\cal A}^{-1/2}e^{i{\bf G}\cdot {\bf r}}$
satisfy the orthonormality condition

\begin{equation}
\frac{1}{\cal A} \int_{\cal A} d^2{\bf r} \,
e^{-i ({\bf G -G'}) \cdot {\bf r}}
= \delta_{{{\bf G}}{{\bf G}'}}~,
%\int_{\cal A} d^2{\bf r} \varphi_{{\bf G}}^{\star} \varphi_{{\bf G}'} =
%\delta_{{{\bf G}}{{\bf G}'}}~,
\end{equation}
where ${\cal A} = a_xa_y$ is the area of a unit cell.
Expanding the product $(\psi_0 f)$ in a similar way, {\em viz.,}

\begin{eqnarray}
(\psi_0 f) 
&=& \psi_0 \delta v_{\rm eff} +\frac{\lambda_w}{2}
\frac{\nabla^2 \psi_0}{\psi_0}  \delta \psi - 
\frac{\lambda_w}{2} \nabla^2 \delta \psi \nonumber \\
&=& e^{i({\bf q} \cdot {\bf r} -\omega t)} \sum_{{\bf G}} 
f_{\bf G} e^{i {\bf G} \cdot {\bf r}}\,,
\label{psi_0f}
\end{eqnarray}
and substituting these expansions into Eq.~(\ref{master}), we obtain
the nonlinear eigenvalue equation

\begin{equation}
\omega (\omega^2 - \omega_{c}^{2}) c_{{\bf G}}
= \omega \sum_{{\bf G}'} B_{{{\bf G}}{{\bf G}'}} f_{{\bf G}'} +
\omega_c \sum_{{\bf G}'} A_{{{\bf G}}{{\bf G}'}} f_{{\bf G}'}~,
\label{eigenvalue}
\end{equation}
where

\begin{equation}
B_{{{\bf G}}{{\bf G}'}} \equiv \frac{1}{2} \left ( {\bf q} + {\bf G}
\right )^2\delta_{{{\bf G}}{{\bf G}'}} + \frac{1}{\lambda_w}
(v_{\rm eff} - \mu) [{\bf G} - {\bf G}']~,
\label{bmat}
\end{equation}
and
\begin{equation}
 A_{{{\bf G}}{{\bf G}'}} \equiv i[{\bf q} \times ({\bf G} - {\bf G}') +
({\bf G} \times {\bf G}')]\cdot \hat{\bf z}~
{\rm ln}\psi_0 [{\bf G} - {\bf G}']~.
\label{amat}
\end{equation}
We adopt the notation that square braces denote a Fourier-transformed
variable.

The Fourier coefficients $f_{{\bf G}}$ are connected to the
fluctuating wavefunction expansion coefficients $c_{{\bf G}}$ through
Eq.~(\ref{psi_0f}), {\em viz.,}
\begin{eqnarray}
f_{{\bf G}} &=& \sum_{{\bf G}'}[M^{K}_{{{\bf G}}{{\bf G}'}} +
M^{X}_{{{\bf G}}{{\bf G}'}} + M^{H}_{{{\bf G}}{{\bf G}'}} +
\lambda_w B_{{{\bf G}}{{\bf G}'}}] c_{{\bf G}'} \nonumber \\
&\equiv& \sum_{{\bf G}'} \tilde{M}_{{{\bf G}}{{\bf G}'}} c_{{\bf G}'}~,
\label{ftoc}
\end{eqnarray}
and
\begin{eqnarray}
M^{K}_{{{\bf G}}{{\bf G}'}} &=& 2\pi \psi_0^{2} [{\bf G} - 
{\bf G}']\nonumber \\
M^{X}_{{{\bf G}}{{\bf G}'}} &=& -\sqrt{\frac{8}{\pi}} 
\psi_0 [{\bf G} - {\bf G}']\nonumber \\
M^{H}_{{{\bf G}}{{\bf G}'}} &=& 4\pi \sum_{{\bf G}''}
\frac{\psi_0 [{\bf G} - {\bf G}'']\psi_0 [{\bf G}'' - {\bf G}']}
{\sqrt{(q_x + G_x'')^2 + (q_y + G_y'')^2 }}~,
\label{matrices}
\end{eqnarray}
are the kinetic, exchange, and Hartree matrices that
arise when the $\psi_0 \delta v_{\rm eff}$ term is Fourier transformed.
The eigenvalues $\omega$ of Eq.~(\ref{eigenvalue}) give 
the excitation frequencies of the system, and the corresponding
eigenvectors $\vec{c}$ can be used to
determine the density fluctuation of the collective mode.

Eq.~(\ref{eigenvalue})
represents an infinite dimensional matrix problem that, for practical
purposes, must be solved on some subset of the 2D reciprocal lattice
vectors.  Any truncation of the basis set must be checked to ensure 
that the results for the modes of interest do not depend on the
number of ${\bf G}$ vectors retained.  If the system is weakly 
modulated, the coupling between different reciprocal lattice vectors 
is likewise weak, and the matrix problem is only of a modest size.  
However, if the system is strongly
modulated with a large unit cell $(a \gtrsim 800$ nm), the number of
basis functions required to adequately
describe the dynamics of the system can become unmanageably large.
This is a possible limitation of the plane-wave expansion technique.

Since our main interest is in making contact with the experimental FIR 
data, we shall only consider the ${\bf q}=0$ limit of our 
calculations.\cite{footnote2}
In this situation, our dynamical equations are invariant with respect 
to the point group $C_{4v}$ (for $a_x = a_y$) when $\omega_c =0$.  A 
nonzero magnetic field lowers the symmetry of the system to $C_4$ 
because of the additional term involving
the $A_{{\bf G}{\bf G}'}$ matrix in Eq.~(\ref{eigenvalue}).  
In either case, we can appeal to the 
inherent symmetry of the reciprocal space to substantially reduce the 
number of effective basis vectors considered in a calculation.  
The way in which this is done is discussed in Appendix \ref{appendixa}.

\subsection{Power Absorption}
\label{power}

We now consider the response of the system to a spatially uniform
radiation field incident normally on the sample and polarized in
the $x$-direction:
${\bf E}^{\rm ext} ({\bf r}, t) =\frac{1}{2}E_{0}(e^{-i\omega t} + 
e^{i \omega t}) {\bf \hat x}$.  This is the most direct way of making
contact with experiments which probe the collective excitations of the
system by means of FIR spectroscopy.  

The physically relevant quantity is the time averaged power absorption,
which is given by\cite{vanzyl99}

\begin{eqnarray}
\langle P \rangle_{t} &=& 
\left \langle \int d^2{\bf r}~{\bf j}^{\rm ind}({\bf r},t)\cdot
{\bf E}^{\rm ext}({\bf r},t) \right \rangle_{t} \nonumber \\
&=& \frac{1}{2}
E_{0}~{\rm Re}\int d^2{\bf r}~j_x^{\rm ind} ({\bf r},\omega)~,
\label{avgpwr}
\end{eqnarray}
where $j_x^{\rm ind} ({\bf r},\omega)=- n_0 ({\bf r}) v_x ({\bf r},w)$ 
is the induced current density and the velocity is now a solution of 

\begin{equation}
\frac{\partial {\bf v}}{\partial t} =
- \gamma {\bf v} + \delta {\bf F}^{\rm tot} -
{\bf v} \times {\bf B}^{\rm ext}~.
\label{vel_gamma}
\end{equation}
The total fluctuating force includes the external contribution
$\delta {\bf F}^{\rm ext} = -{\bf E}^{\rm ext}$.
In addition, we have introduced a phenomenological relaxation
rate $\gamma$ which has the effect of giving the excitations a finite 
lifetime. This implies that any frequency factors arising from 
the time-derivative of the velocity must be replaced by 
$\omega \rightarrow \tilde{\omega} = \omega + i \gamma$. 
In previous studies \cite{ye-zaremba1994} 
it was found that the experimental transmission data was most faithfully
reproduced by a frequency dependent relaxation rate, $\gamma = 
\gamma_0/\omega$. We retain the same frequency dependence here.

Noting that the current density is itself a periodic function of
both spatial directions, we can write Eq.~(\ref{avgpwr}) as

\begin{equation}
{\langle P \rangle_{t} \over A} = \frac{1}{2} E_{0}~{\rm Re}~
j_{x}^{\rm ind}[ {\bf G}=0,\omega]~,
\label{pwr_g}
\end{equation}
where $A$ is the sample area and
\begin{equation}
j_{x}^{\rm ind}[{\bf G},\omega] = \frac{1}{\cal A} \int d^2{\bf r}~ 
e^{-i{\bf G}\cdot {\bf r}} j_{x}^{\rm ind}({\bf r},\omega)
\label{jg}
\end{equation}
is the Fourier coefficient of the induced current. The ${\bf G} =0$
component is just the average induced current within a unit cell.
A straightforward calculation paralleling Ref. [\onlinecite{vanzyl99}] 
leads to the following expression for this component:

\begin{eqnarray}
j_{x}^{\rm ind}[{\bf G}=0,\omega] &=& \frac{-2 \tilde{\omega}}{\tilde{\omega}^2 - \omega_c^2}
\sum_{{\bf G}'} G_x' \psi_0 [{\bf G}'] f_{{\bf G}'} \nonumber \\ &+&
\frac{2 i \omega_c }{\tilde{\omega}^2 - \omega_c^2}
\sum_{{\bf G}'} G_y' \psi_0 [{\bf G}'] f_{{\bf G}'} 
+ \frac{i\tilde{\omega} n_0[{\bf G}=0]E_0}{\tilde{\omega}^2 - \omega_c^2}~.
\label{j_G=0}
\end{eqnarray}
The net effect of including an external driving
field ${\bf E}^{\rm ext}$ was previously shown to convert the 
nonlinear eigenvalue problem in Eq.~(\ref{eigenvalue}), into a set of 
inhomogeneous equations\cite{vanzyl99}

\begin{equation}
\omega (\tilde{\omega}^2 - \omega_{c}^{2}) c_{{\bf G}}
- \tilde{\omega} \sum_{{\bf G}'} B_{{{\bf G}}{{\bf G}'}} f_{{\bf G}'} -
\omega_c \sum_{{\bf G}'} A_{{{\bf G}}{{\bf G}'}} f_{{\bf G}'} = 
b_{\bf G}~,
\label{inhomoeigenvalue}
\end{equation}
where

\begin{equation}
b_{\bf G} = -E_0 (i\tilde{\omega}G_x -\omega_c G_y ) \psi_0 [{\bf G}]~.
\label{inhomovec}
\end{equation}
The ${\bf q} = 0$ solution of Eq.~(\ref{inhomoeigenvalue}) along with
Eq.~(\ref{ftoc}) provides a complete determination of
$j_{x}^{\rm ind}[{\bf G}=0,\omega]$.  The substitution of
Eq.~(\ref{j_G=0}) into Eq.~(\ref{pwr_g})
yields the final expression for the FIR power absorption.

\section{QUANTUM DOTS}
\label{quantumdots}

In Sec.~\ref{tfdw_formalism}, the formalism for studying a 2D
modulated electronic system was developed for an arbitrary 
periodic confining potential $v_{\rm ext}({\bf r})$.
In this section, we restrict our attention to potentials that
confine the electrons in all three spatial dimensions, thereby
forming an array of quasi-0D quantum dots.
In particular, we have in mind the experiment performed by Demel
{\em et al.}~\cite{demel1990,demel1990b}  in which an array of quantum 
dots was prepared from  modulation--doped
AlGaAs/GaAs heterostructures.  In this experiment, resonant 
anticrossings were observed in the magnetic dispersions 
for arrays of dots with $N=25$ and $N=210$ electrons per dot.
These anticrossings were tentatively explained in terms of a nonlocal 
interaction associated with the finite compressibility of the electron 
gas,~\cite{demel1990,demel1990b} although subsequent work pointed to
the importance of noncircular perturbations. Here we explore the 
origin of these anticrossings in greater detail by systematically 
studying confining potentials of the form

\begin{equation}
v_{\rm ext}(r,\theta) = \frac{1}{2}kr^2 + \frac{1}{4} p r^4 +
\frac{1}{4} p \varepsilon r^4 {\rm cos}(4\theta)~,
\label{confpot}
\end{equation}
where $\theta$ is the azimuthal angle in polar coordinates. 
For $p=0$, we have parabolic confinement with $\omega_0 = \sqrt{k}$.
Setting $\varepsilon =0$ with $k,p\neq 0$ introduces a
quartic radial perturbation, and
finally, if all three parameters are nonzero, the last term in 
(\ref{confpot}) breaks the circular symmetry of the
potential. It is understood that the potential in Eq.~(\ref{confpot}) is
centered in the unit cell and periodically extended throughout the 
array.  Since the dots of interest are localized near the center of 
the cell, the behaviour of the potential near the cell boundaries is 
of no consequence.

In order to make contact with previous work on this problem, it will 
prove useful to elaborate on the various parameters
appearing in Eq.~(\ref{confpot}). To begin, we set $\varepsilon = 0$.
Following Ref.~[\onlinecite{ye-zaremba1994}], we define
the anharmonicity parameter $\eta \equiv p R_0{^2}/2k$, where $R_0$ is
the radius of the dot as determined within a classical calculation
which includes only the Hartree interaction of the electrons.
This radius is given implicitly by the equation~\cite{ye-zaremba1994}

\begin{equation}
R_0 = \frac{R_0^{\rm p}}{(1+\frac{8}{5}\eta)^{1/3}}~,
\label{R_0}
\end{equation}
where $R_0^{\rm p}$ is the equilibrium radius for pure parabolic 
confinement (hence the superscript `p') given by~\cite{nazin1989}

\begin{equation}
R_0^{\rm p} = 
\left [ \frac{3\pi}{4} \frac{e^2N}{\epsilon k} \right ]^{1/3}.
\label{R_0p}
\end{equation}
It was further shown in Ref.~[\onlinecite{ye-zaremba1994}] 
that the zero-field ($B = 0$) frequency of the lowest dipolar mode 
is given to an excellent approximation by

\begin{equation}
\omega_{0}' = \omega_0\sqrt{1 + \frac{8}{5}\eta}~,
\label{w0p}
\end{equation}
where $\omega_0$ is the bare harmonic frequency defined above.
Thus, in the absence of a noncircular perturbation, a given zero-field
dipole frequency is determined by a family of $\omega_0$ and $\eta$
values. These values are adjusted to reproduce the experimentally
observed frequency $\omega_0'$, as well as other features of the
observed absorption spectrum. In this way, the quantities $k$ and $p$
can be determined. We observe from our calculations that the subsequent
inclusion of the
parameter $\varepsilon$ has a negligible effect on $\omega_0'(B=0)$, so
that the effect of this parameter can be investigated independently of
the other two.  Typical values for $\eta$ and $\varepsilon$ used in the
calculations lie in the range $[0,1]$.

Finally, we summarize the appropriate physical parameters for 
the AlGaAs/GaAs heterostructures under consideration:
the effective Bohr radius $a_0^{\star}$ is {103~\AA} and the effective
Rydberg of energy, Ry$^{\star} = e^2/2\epsilon a_0^{\star}$, is 5.4 meV.
To be as consistent as possible with the experimental
situation,~\cite{demel1990}
we focus on dot arrays with $N=210$ electrons per dot and a FIR
zero field absorption frequency of $\omega_0^{'} = 32~{\rm cm}^{-1}$.

\subsection{Parabolic Confinement}
\label{parabolic}

As mentioned above, $\eta = 0$ provides a potential that defines
an array of parabolically confined dots.  The nature of the collective
excitations for such a system are by now well established in the 
literature.
\cite{nazin1989,maksym1990,shikin1991,gudmundsson1991,pfannkuche1991,nazin1992,ye-zaremba1994,bakshi1990,broido1990,dempsey1990,huang1991,pfannkuche1994,wagner1995,gudmundsson1995,lipparini1997}
In Fig.~\ref{density}(a) we show the calculated 
ground state density for a square array with $a_x = a_y = 1000$ nm
containing well-separated dots with $N=210$ electrons per dot.
The radius of the dot is about 150 nm, and for the sake of clarity, 
we show only that portion of the unit cell in which the electronic 
density is localized. The radius found here on the basis of the TFDW
functional is similar to the value obtained using Eq.~(\ref{R_0p}), 
{\em viz.,} $R_0^{\rm p} = 160$ nm, with $\omega_0 = 32~{\rm cm}^{-1}$ 
and $N=210$. The similarity of these values indicates that the Hartree
interaction is in fact dominating the equilibrium distribution and that
the exchange and kinetic energies are playing a relatively minor role.
The shape of the distribution is approximately semi-circular as found in
the classical calculation~\cite{ye-zaremba1994}, but with a smooth
variation at the edge of the dot as a result of the von Weisz\"acker
correction. This term in the energy functional allows us to simulate 
the behaviour of the charge density expected within a fully 
quantum-mechanical calculation.

It is perhaps worth emphasizing that the equilibrium calculations
described above effectively correspond to {\it isolated} quantum dots.
The Coulomb interactions between equilibrium charge densities in 
different cells of the array are eliminated~\cite{}, so that only
intra-cell Coulomb interactions are retained in the calculation. Of
course the potential within a given cell arises from all other
charges in the system, and one should view the parabolic confining
potential as the net effect of all these other charges. However, when
calculating the magnetoplasmons, the inter-cell Coulomb interactions 
arising from the density fluctuations are {\it not} neglected. These
multipolar interactions are required in order to simulate the response
of an array of coupled dots. We shall examine shortly the importance of 
these inter-dot Coulomb interactions. 

In Fig.~\ref{B-paradot}, we show the calculated magnetic dispersion 
for the array of dots described above.  To avoid cluttering the diagram,
we have chosen to omit the large number of edge modes which start from 
$\omega = 0$ at $B=0$ and lie below $\omega = \omega_c$. These modes 
make no observable contribution to the power absorption and are 
therefore of little interest in the present paper. The mode frequencies
found here are very similar to those found in the classical
approximation~\cite{nazin1989} and can be classified according to the 
local $SO(2)$ symmetry of circularly symmetric confining potentials. 
As discussed in Appendix \ref{appendixa}, the Lie group $SO(2)$ has 
$m$ one-dimensional irreducible representations given by 
$\chi^m(\phi) = e^{im\phi}$, with $0 \le \phi \le 2\pi$, and $m \in Z$.
The different branches seen crossing in Fig.~\ref{B-paradot} correspond
to different irreducible representations of $SO(2)$ (see Table 
\ref{charso2} in Appendix \ref{appendixa}), and thus one would not 
expect any mode repulsion in the absence of some perturbation.
%To confirm this, we show a closeup view of the boxed region containing
%the crossing between the $1^{+}$ and $3^{-}$ modes. Even on the 
%scale of this figure, the modes pass cleanly through each other.
Since the only possible perturbation in the present
situation is the dynamic inter-cell Coulomb interaction, the results
shown here confirm that this interaction is very weak. As we shall see,
the effect of interdot interactions can be made apparent by
reducing the interdot separation.

Within the classical theory, a subset of the dot modes has a dispersion 
given by the expression~\cite{nazin1989,shikin1991,ye-zaremba1994}

\begin{equation}
\omega_{j,\pm} = \sqrt{\Omega_j^2 + (\omega_c/2)^2} \pm \omega_c/2~.
\label{wpm}
\end{equation}
The solid curves in Fig.~\ref{B-paradot} show some fits to 
Eq.~(\ref{wpm}), and demonstrate that the field dependence is well
reproduced even though the TFDW values of $\Omega_j$ are slightly
different from the analytic classical values. The lowest pair of curves
corresponds to the two circular polarizations of the center-of-mass 
(CM) mode. According to the generalized Kohn theorem,\cite{dobson1994} 
the exact 
separability of the $N$-body Hamiltonian into CM and relative 
coordinates for parabolic confinement ensures that $\Omega_1 = \omega_0$
($\omega_0 = 32$ cm$^{-1}$ in this case), and that only these dipole
modes couple to a uniform external electric field.  This fact is 
illustrated in the inset to Fig.~\ref{B-paradot} where we show
the calculated FIR power absorption for the dot array for a range of
magnetic fields consistent with the experiment in 
Ref.~[\onlinecite{demel1990}].
The single peak at $\omega(B=0)=32~{\rm cm}^{-1}$ reflects the fact 
that our model satisfies the generalized Kohn theorem.  It is also clear
from this figure that the two-peak structure appearing for $B\neq 0$,
corresponds to the CM-mode dispersion seen in Fig.~\ref{B-paradot}.

We next consider the effects of interdot Coulomb coupling on the
magnetoplasmon excitations. We have already indicated that this
interaction is weak for a lattice constant of $a = 1000$ nm, and we now
support this assertion by observing the effect of reducing the lattice
constant to $a = 600$ nm. Fig.~\ref{B-closeparadot} shows a small
section of the magnetic dispersion for this case.  Although this 
dispersion looks similar to Fig.~\ref{B-paradot}, there are some 
important differences.  

First, we notice a slight red-shift of the 
$\omega_{0}(B=0)$ frequency while the higher-lying modes have moved
slightly in the opposite direction. The softening of the 
$\omega_{0}(B=0)$ frequency can be explained by invoking the interaction
between the induced dipoles on each lattice site.
For the square lattice being considered, the electric field experienced
by the $m$-th dipole due to all other dipoles is given
by~\cite{jackson}

\begin{eqnarray}
E_m =  \frac{1}{a^3} {\sum_{l,n}}' \frac{1}{(l^2 + n^2)^{3/2}} \left [
\frac{3 l^2}{l^2 + n^2} -1 \right ] p_{m+l},
\label{E_m}
\end{eqnarray}
where we have assumed, consistent with Sec. \ref{power}, that 
the polarization is in the $\hat{\bf x}$-direction. 
The quantities $l$ and $n$ identify positions on the 2D lattice, 
{\em viz.,} ${\bf r} =  l a\hat{\bf x} + n a\hat{\bf y}$, and the 
primed summation indicates that the $n=l=0$ point is excluded. In the 
FIR regime, each of the dipole moments $p_{m}$ has a common magnitude
$p_0$, and the electric field can be written as

\begin{equation}
E_m = \frac{p_0}{a^3} S~,
\label{E_m2}
\end{equation}
where
\begin{equation}
S = {\sum_{l,n}}' \frac{1}{(l^2 + n^2)^{3/2}} \left [
\frac{3 l^2}{l^2 + n^2} -1 \right ]~.
\end{equation}
The quantities $p_{m}$ and $E_m$ are connected through the
relation $p_{m} = \alpha(\omega) E_m$ where
$\alpha(\omega)$ is the dipole polarizability of the dot.
For parabolic confinement, the polarizability is given by
$\alpha (\omega) = Ne^2/m^{\star}(\omega_0^2 - \omega^2)$. 
Using this result in
Eq.~(\ref{E_m2}), we find the following frequency for the CM mode:

\begin{eqnarray}
\omega^2 &=& \omega_0^2 - \frac{S}{4\pi}\omega_p^2~, 
\label{CM-dispersion}
\end{eqnarray}
where $\omega_{p}^{2} = 4\pi Ne^2/\epsilon m^{\star} a^3$ is the 
effective plasmon frequency.
This simple result shows that the interdot dipole interactions
decrease the absorption frequency from that of a single dot.  

Next, we note the appearance of small anticrossings in the magnetic 
dispersion that was not discernible for the well separated dots.
These anticrossings are still difficult to resolve, so
in the figure inset, we show a close-up view of the boxed region which
clearly illustrates the mode repulsion.  These anticrossings
arise because the interdot coupling of the dots on the square lattice 
breaks the {\em local} circular symmetry of the confining potential.
Specifically, the square symmetry of the lattice is lower than the 
local $SO(2)$ symmetry of the dots. As shown in 
Appendix~\ref{appendixa}, the $1^{+}$ and $3^{-}$ modes now transform 
under the same irreducible representation
of $C_4$ (see Table \ref{charc4} in Appendix \ref{appendixa}),
which means that the modes will repel (anticross) each other under 
a reduction of the symmetry.  We stress however that these 
anticrossings are far too weak to account for the mode anticrossings 
observed in the Demel {\em et al.} experiments.

\subsection{Radial Anharmonic Confinement}
\label{r4}

% talk about the splitting in the power absorption and the similarities
% to the parabolic case because of the non interacting mode crossings

The inclusion of a perturbative term of the form 
$\delta v_{\rm ext}(r) = {1 \over 4}pr^4$ in Eq.~(\ref{confpot}) 
preserves the circular symmetry of the 
confining potential, but allows for the mixing of different dipole
modes with a redistribution of the FIR oscillator strength.
The ground state density is illustrated in Fig.~\ref{density}(b) 
and is qualitatively similar to that found in the classical 
model.~\cite{ye-zaremba1994} This density was generated with
$\omega_0 = 20 {\rm cm}^{-1}$ and $\eta = 1$. The circular ridge is a 
result of the $r^4$ perturbation. The calculated magnetic dispersion 
and FIR power absorption are shown in Fig.~\ref{B-r4dot}.

Comparing Figs.~\ref{B-paradot} and \ref{B-r4dot}, we
see that many of the features of the magnetic dispersion in the case of
parabolic confinement are still present for the anharmonic potential.  
To emphasize this point, we have plotted the dispersions described by 
Eq.~(\ref{wpm}) using $\Omega_j$ as an adjustable parameter.
We see that Eq.~(\ref{wpm}) provides a good fit to the data, and
indicates that the functional form of the magnetic dispersion is not
sensitive to the form of the confining potential. In particular, a fit
of this dispersion to the lowest CM modes does {\it not} in general
provide a direct measurement of the harmonic term in the confining
potential. The relative spacing between the CM modes and the higher
modes does depend on the anharmonicity, but the latter are not usually
observable in FIR experiments. 

The main observable difference between parabolic and radially anharmonic
confinement is revealed by the theoretical power absorption in the inset
to Fig.~\ref{B-r4dot}. The curves were generated for a
range of magnetic fields corresponding to the experimental situation
in Ref.~[~\onlinecite{demel1990}]. The dominant dipolar peaks found for
parabolic confinement are now accompanied by a weak satellite which
tracks along the $\omega_+$ CM-like mode. The oscillator strength of the
satellite is directly controlled by the parameter $\eta$. A value of
approximately 1 reproduces the observed oscillator strength at high
fields, while the harmonic confining frequency was adjusted to $\omega_0
= 20$ cm$^{-1}$ in order to yield a $B=0$ dipole frequency of 32
cm$^{-1}$. These parameters imply a strong anharmonicity. The 
$B=0$ peak at $\omega = 32~{\rm cm}^{-1}$ in Fig.~\ref{B-r4dot}
can therefore no longer be identified as a CM mode in the sense of
parabolic confinement. The very appearance of the satellite is
an indication that the generalized Kohn theorem is no longer valid.
The $r^4$ perturbation has the effect of coupling the dipolar modes
found in the parabolic limit,~\cite{ye-zaremba1994} thereby making
other dipole modes FIR active.  A comparison of our calculated FIR 
power absorption to the experimental data~\cite{demel1990}
leaves little doubt that 
the satellite structure observed in the experiment has its origins in 
the anharmonicity of the confining potential. This conclusion 
confirms that reached by other workers using different theoretical 
approaches.~\cite{gudmundsson1991,pfannkuche1991,ye-zaremba1994}
The power absorption thus provides a direct probe for determining
the geometry of the confining potential.

Finally we mention that calculations were also carried out for the
reduced lattice spacing of 600 nm. As in the case of parabolic
confinement, anticrossings in the magnetic dispersion arise, but the
strength of these anticrossings is again too weak to account for those
observed experimentally. We can therefore rule out interdot Coulomb
interactions as a significant mechanism at the experimental interdot
separations.

\subsection{Noncircular Confinement}
\label{noncircular}

We now consider an explicit symmetry breaking perturbation of the form
given in Eq.~(\ref{confpot}) with $\varepsilon \neq 0$. It is clear
that such a perturbation is consistent with the geometry of the dots
studied in Ref.~[\onlinecite{demel1990}].  The equilibrium 
density for the noncircular confining potential is shown in
Fig.~\ref{density}(c).  This density profile was generated with
$N=210$, $\omega_0 = 20~{\rm cm}^{-1}$, $\eta = 1$, $\varepsilon = 0.4$,
and $a_x = a_y = 1000$ nm. The maximal radial extent of this dot is 
about 155 nm.
It is clear from Fig.~\ref{density}(c) 
that the deviation from circular symmetry is very pronounced,
and one should not expect the same mode dispersions as obtained for
arrays of circularly symmetric dots.

Turning to the magnetic dispersion in Fig.~\ref{B-noncirc},
we do indeed find a very different $B$-field dependence of the
collective modes.  Specifically, we note the appearance of strong 
anticrossings between the lowest-lying modes, and progressively weaker
anticrossings for the higher modes. These anticrossing are a 
consequence of the $\delta v_{\rm ext}(r,\theta)={1\over 4}p 
\varepsilon r^4 {\rm cos}(4\theta)$ perturbation in Eq.~(\ref{confpot}),
and occur when the angular momentum of the symmetric modes differ by
an integral multiple of four. The lowest anticrossing is between a pair
of $m=1$ and $m=-3$ modes and is the most important since it is the 
structure observed in the Demel {\em et al.} experiment. A comparison 
of Fig.~\ref{B-noncirc} with Fig.~2 of Ref.~[~\onlinecite{demel1990}] 
shows that our results are in good agreement 
with regard to the size and location of the anticrossing. 
Once again, the solid curves in Fig.~\ref{B-noncirc} are fits to 
Eq.~(\ref{wpm}) and here, they serve to emphasize the difference 
between circular and noncircular confinement geometries.  

The magnitude of the gap occurring at 
each anticrossing is directly related to the parameter $\varepsilon$.  
The value $\varepsilon= 0.4$ was chosen to best fit the observed
transmission data~\cite{demel1990} and the inset to
Fig.~\ref{B-noncirc} shows our calculated power absorption.
At low fields there is weak structure to the high field side of the main
resonance which is probably too weak to be resolved experimentally.
However, as one enters the field range of the anticrossing between 1 and
2 T, a new peak appears and leads to an absorption spectrum consisting
of three peaks. This structure is most distinct at a field of $B=1.5$ T
which is in the middle of the anticrossing region. By $B=2.4$T, the 
central peak has already lost most of its oscillator strength, and 
the power absorption takes on the characteristics of the $r^4$ power
spectrum of Fig.~\ref{B-r4dot}, including the high-field satellite.
This overall behaviour is entirely consistent with 
experiment.~\cite{demel1990} 
  
\section{CONCLUSIONS}
\label{conclusions}

In this paper, we have presented a generalization of our previous work 
\cite{vanzyl99} on TFDW hydrodynamics in laterally modulated electronic
systems. As a specific application, we have examined the magnetoplasmon 
excitations in arrays of both circular and noncircular dots. By 
considering potentials which include both radial and noncircular 
anharmonic perturbations, we are able to fully explore the effects of
geometric confinement on the magnetoplasmon excitations of the dots.

In the case of circular dots, our results indicate that an $r^4$ radial
perturbation can account for the satellite peak structure observed in 
the experiments.\cite{demel1990,demel1990b,ye-zaremba1994} We have also
seen that the interdot coupling is too weak at the experimental lattice
constant to give rise to any discernible anticrossing effects, in
agreement with earlier results.\cite{bakshi1990}
These effects only become apparent when the lattice spacing is reduced
to the order of the size of a dot.

On the other hand, the addition of a noncircular perturbation 
of the form $r^4{\rm cos}(4\theta)$ was shown 
to induce anticrossings in the
magnetoplasmon dispersion that are entirely consistent with  
the experimentally observed transmission data.  Specifically, the
location for the onset of the first optically observable anticrossing,
along with higher $B$-field peak structure, was found to be in good 
agreement with experiment. Based on this result, we conclude that only 
an explicit noncircular symmetry in the confining potential can fully 
account for the mode anticrossings observed in the 
experiments.\cite{demel1990,demel1990b}

The work presented here is applicable to a wide variety of 2D
geometrical confinements.  Our current interest lies in the application
of this formalism to antidot arrays which are complementary structures
to quantum dot arrays. Recent experimental work on these 
systems has revealed a collective excitation spectrum 
very different from the dot arrays.\cite{kern1991,zhao1992}  
A detailed discussion of these systems will be presented elsewhere.

\acknowledgments
This work was supported by a grant from the Natural Sciences and
Engineering Research Council of Canada.  

\appendix
\section{DERIVATION OF THE SYMMETRY-REDUCED DYNAMICAL EQUATIONS}
\label{appendixa}

The square primitive cell of our problem results in
a reciprocal lattice with 4-fold symmetry, as illustrated in 
Fig.~\ref{appfig1}. The point group of the square lattice
is $C_{4v}$, and the full point symmetry is $C_{4v}\otimes(E,T)$, where
$(E,T)$ is the time-reversal symmetry group and $E$ is the identity 
element.~\cite{grouptheory} 
In the context of our problem, we recall that we are primarily
interested in the calculation of the FIR response of the system, 
which corresponds to the ${\bf q} = 0$ limit of the general set of 
inhomogeneous equations, Eq.~(\ref{inhomoeigenvalue}).
In the absence of a magnetic field, these equations are 
invariant under $C_{4v}\otimes(E,T)$,
however the inclusion of a magnetic field breaks time-reversal
symmetry and lowers the symmetry of the system from $C_{4v} 
\rightarrow C_4$.  This is easily seen if one observes the symmetry 
group of the $A_{{\bf G}{\bf G}'}$ matrix in Eq.~(\ref{amat}).
It follows that $C_4$ is the relevant group symmetry to consider in
the most general case (i.e., magnetoplasma excitations).

Let us step back for a moment, and consider the full symmetry group of 
the square lattice, namely, $C_{4v} = \{E,C_4,C_4^3,C_4^2,m_x,m_y,
\sigma_u, \sigma_v\}$, where $m$ are reflections about the $x$ or $y$ 
axes, and $\sigma$ is a reflection about a diagonal of the square.  
We can construct a representation, $\Gamma$ (we use this notation to 
remind us that we are at the $\Gamma$ point of the first BZ), of this 
group by considering a function space spanned by the set of functions 
$\{|{\bf G}_i\rangle\}$.  One natural choice would be to consider the 
set of functions defined by $|{\bf G}_i\rangle \equiv a^{-1} 
e^{i {\bf G}_{i} \cdot {\bf r}}$ with $1 \le i \le 4$ as shown in 
Fig.~\ref{appfig1}.  If we choose as our
canonical function $|{\bf G}_1\rangle$, the other 
functions are obtained by acting with the $C_4$ operator 
on the  function $|{\bf G}_1\rangle$: 
$C_4^n$ $|{\bf G}_1\rangle$ = $|{\bf G}_{1+n}\rangle$, with
$ |{\bf G}_{ 5} \rangle \equiv  |{\bf G}_{1} \rangle$.
Since the generators of the group $C_{4v}$ are $\{ C_4,m_x\}$, we need 
only obtain the representations for the generators of $C_{4v}$ to 
construct the entire representation of the group.  
It is easy to show that generators of $C_{4v}$ in this 
representation take the form

\begin{eqnarray}
\Gamma(C_4)  =
\left[ \begin{array}{rrrr}
0&1&0&0\\
0&0&1&0\\
0&0&0&1\\
1&0&0&0\\
\end{array} \right ]~~~~~~~~~~~~~~~ 
\Gamma(m_x) = 
 \left[ \begin{array}{rrrr}
0&0&0&1\\
0&0&1&0\\
0&1&0&0\\
1&0&0&0\\
\end{array} \right ]~,
\end{eqnarray}
and that the characters for $C_{4v}$ are given by

\begin{center}
\begin{tabular}{ccccccccccccc}
&$C_{4v}$& &$E$& &$2C_4$& &$C_4^2$& &$2m$& &$2\sigma$&\\ \hline
&$\Gamma$& &4& &0& &0& &0& &2&\\
\end{tabular}
\end{center}
The character table for the irreducible representation of $C_{4v}$ is 
shown in Table \ref{charc4v}.  Since there is no
four-dimensional irreducible representation of $C_{4v}$, $\Gamma$ must
be reducible.  It is a simple matter of class-wise character addition to
determine that the $\Gamma$ representation must be decomposed according
to $\Gamma = \Gamma_1 \oplus \Gamma_4 \oplus \Gamma_5$.

Now, let us consider the dipole operator ${\bbox \mu} = e{\bf r}$.  This
operator generates the representation 
$\Gamma_{{\bbox \mu}} = \Gamma_1 \oplus \Gamma_5$ under $C_{4v}$. The 
$\Gamma_1$ representation is derived from the $z$-component of 
${\bf r}$, whereas the $\Gamma_5$ representation is induced by the 
$(x,y)$ components. Owing to the fact that we are only considering 
radiation polarized in the plane of the 2DEG, we can immediately see 
that the invariant subspace of $\Gamma_5$ contains all of the dipole 
active modes of our problem. Thus, rather than concerning ourselves 
with solving the generalized set
of equations, {\it viz.,} Eq.~(\ref{inhomoeigenvalue}), we can use the
symmetry of the system to solve for only those eigenvalues that will be
of interest in a FIR mode calculation.

The problem of projecting into the $\Gamma_5$ representation is really a matter
of block-diagonalizing our system of equations and picking out that block 
associated with $\Gamma_5$.  This amounts to finding a unitary transformation
that will block-diagonalize each of the $\Gamma$ matrices of the group 
$C_{4v}$.
Motivated by the knowledge that under a nonzero magnetic field the group
symmetry is $C_4$, we consider the eigenvalues and eigenvectors of the
group element $C_4 \in C_4$.  Indeed, since $C_4$ is the generator of the
point group $C_4$, it is the only group element that we need to consider.
A simple calculation reveals that the eigenvalues and normalized eigenvectors
of $\Gamma(C_4)$ are 

\begin{eqnarray}
\lambda^{(1)} &=& 1~~~~~~|v^{(1)}\rangle = \frac{1}{2}\left [\begin{array}{r}
1\\1\\1\\1 \end{array} \right ]~;
~~~~~~\lambda^{(2)}= -1~~~~~~|v^{(2)}\rangle = \frac{1}{2}\left [\begin{array}{r}
1\\-1\\1\\-1 \end{array} \right ] \nonumber \\
\lambda^{(3)} &=& i~~~~~~|v^{(3)}\rangle = \frac{1}{2}\left [\begin{array}{r}
1\\i\\-1\\-i \end{array} \right ];~~~~~~
\lambda^{(4)} = -i~~~~~~|v^{(4)}\rangle = \frac{1}{2}\left [\begin{array}{r}
1\\-i\\-1\\i \end{array} \right ]~.
\end{eqnarray}
The four normalized eigenvectors $|v^{(\lambda)}\rangle$ are
applied in this order to produce the unitary matrix $U$, and its
inverse $U^{-1}$

\begin{eqnarray}
U=\frac{1}{2} \left [\begin{array}{rrrr}
1&1&1&1\\
1&-1&i&-i\\
1&1&-1&-1\\
1&-1&-i&i\\
\end{array}
\right ]~~~~~~~~~~~~~~~
U^{-1} =\frac{1}{2} \left [ \begin{array}{rrrr}
1&1&1&1\\
1&-1&1&-1\\
1&-i&-1&i\\
1&i&-1&-i\\
\end{array}
\right ] ~.
\end{eqnarray}
For example, applying $U$ to the group element $C_4$ yields

\begin{eqnarray}
U^{-1}C_4 U &=& \left [ \begin{array}{rrrr}
1&0&0&0\\
0&-1&0&0\\
0&0&i&0\\
0&0&0&-i\\
\end{array}
\right ] ~.
\end{eqnarray}
The unitary transformation matrix, $U$, will block-diagonalize all of the
$\Gamma$ matrices for the group $C_{4v}$.  In other words, the eigenvectors
$\{|v\rangle\}$ of $C_4$ define a symmetrized basis for $\Gamma$ of
$C_{4v}$.  For the general case, $\omega_c \neq 0$, the symmetry is
$C_4$, and all the matrices in the $\Gamma$ representation of $C_4$ will consist
of $1 \times 1$ blocks (i.e., fully diagonalized).
It is now clear that the eigenvalues $\lambda^{(1)}$ and
$\lambda^{(2)}$ are associated with the $\Gamma_1$ and $\Gamma_4$
irreducible representations
of $C_{4v}$ respectively.  The eigenvalues 
$\lambda^{(3)}$ and $\lambda^{(4)}$
are associated with $\Gamma_5$; the vector space spanned by 
their eigenvectors $(|v^{(3)}\rangle,|v^{(4)}\rangle)$ is the two-dimensional
irreducible representation of $\Gamma_5$.

It is of interest to know how the irreducible representations of
$C_{4v}$ relate to those of $C_4$. 
This can be done by appealing to the character tables for both
the $C_{4v}$ and $C_4$ point groups (see Tables \ref{charc4v} and 
\ref{charc4}).  By inspection, we can see that 

\begin{eqnarray}
&\Gamma_1& \rightarrow D_1 \nonumber \\
&\Gamma_2& \rightarrow D_2 \nonumber \\
&\Gamma_3& \rightarrow D_1 \nonumber \\
&\Gamma_4& \rightarrow D_2 \nonumber \\
&\Gamma_5& \rightarrow D_3 \oplus D_4 ~.
\label{branchingrules}
\end{eqnarray}
From Eq.~(\ref{branchingrules}), we see that the two-dimensional
irreducible representation
$\Gamma_5$ {\em splits} when the symmetry is lowered from
$C_{4v} \rightarrow C_4$.
Note also that $D_3$ and $D_4$ are complex conjugate 
representations.  The relevance of this fact is that in the absence of a
magnetic field, these two representations are degenerate (time-reversal
symmetry).  The application of a magnetic field will lift this degeneracy,
and one should expect two modes appearing from the $\Gamma_5$ contribution;
one from each of the $D$'s appearing in its decomposition.
In the context of our model, the absence of a magnetic field implies that we
only require {\em one} of either $\lambda^{(3)}$ or $\lambda^{(4)}$ to
obtain all of the FIR-active modes of the system.

It is also of interest to determine what the relations are between
the irreducible representations of $C_4$ and $SO(2)$.  The (compact) Lie
group $SO(2)$ is the symmetry group of the circular symmetric dots.  If the
dots are well isolated, they are unaware of the lattice, and the 
modes of the system can be classified according to their transformation
under $SO(2)$.  If the dots are closer together (close enough to become
aware of neighbouring dots), then the square symmetry of the lattice will
break the local $SO(2)$ symmetry.  In Table \ref{charso2}, we show the
character table for some of the irreducible representation of $SO(2)$ for
the point group $C_4$.  What is of note is the compatibility relations
between the group  $SO(2)$ and $C_4$, {\it viz.,}

\begin{eqnarray}
&0^+& \rightarrow D_1 \nonumber \\
&0^-& \rightarrow D_1 \nonumber \\
&1^+& \rightarrow D_3 \nonumber \\
&1^-& \rightarrow D_4 \nonumber \\
&2^+& \rightarrow D_2 \nonumber \\
&2^-& \rightarrow D_2 \nonumber \\
&3^+& \rightarrow D_4 \nonumber \\
&3^-& \rightarrow D_3 
\label{branchingrules2}
\end{eqnarray}
Notice that if we are restricting our calculations to the invariant subspace
of $\Gamma_5$, we will only be projecting out the modes with {\em odd}
$m$; the even modes are generated by projecting into $\Gamma_1$ and
$\Gamma_4$.
From these compatibility relations, we see that the modes $1^+$ and 
$3^-$ will mix (anticross) under the symmetry lowering from 
$SO(2) \rightarrow C_4$ because they both transform under the same 
irreducible representation of $C_4$.  
An examination of Table \ref{charso2} reveals that
when the symmetry is strictly $SO(2)$, these modes belong to different 
irreducible representations, and will therefore cross.  The same argument can be used
if there is an explicit symmetry breaking via the confining potential.
Specifically, even if the dots are well separated, a confining potential
that does not transform under $SO(2)$ will lower the symmetry of the 
system and cause mode mixing (anticrossing) to occur.

So far, our discussion has been restricted to an abstract square 
lattice.  How do we apply these ideas to our problem?
We note that our entire reciprocal lattice is made up of 
sets of reciprocal lattice vectors $\{ |{\bf G}_{i\alpha}\rangle \}$,
($\alpha = 1,...,4$), which are generated by the symmetry operation
$C_4$ acting on $|{\bf G}_{i 1}\rangle$. 
It is therefore sufficient to only consider the reciprocal lattice 
vectors within the first quadrant of the total reciprocal space.
The index $i=1,...,n_{max}$ labels these  vectors, and the set 
$\{ |{\bf G}_{i\alpha}\rangle \}$ will be referred to as a {\it shell}.
The total number of shells retained in the calculation is  $n_{max}$.

The Fourier expansion of an 
arbitrary function $f({\bf r})$ can then be written as

\begin{equation}
|f\rangle = \sum_{i \alpha} f_{i \alpha} |{\bf G}_{i \alpha}\rangle~,
\end{equation}
where the amplitudes $f_{i\alpha}$ are the elements of a $4n_{max} 
\times 1$ column vector.  Each set of amplitudes $f_{i\alpha}$ ($\alpha
= 1,..,4$) can be expressed in terms of the eigenvectors of 
$\Gamma(C_4)$ as

\begin{equation}
f_{i\alpha} = \sum_{\lambda} f^{(\lambda)}_i v^{(\lambda)}_\alpha~,
\label{redf}
\end{equation}
and the orthonormality of the eigenvectors implies the inverse relation
\begin{equation}
f_i^{(\lambda)} = \sum_{\alpha} f_{i\alpha} v^{(\lambda)\star}_\alpha~.
\label{inverse}
\end{equation}
The star indicates complex conjugation.
Now, consider some operator ${\cal O}$ which is {\it invariant} under 
$C_4$.  The matrix representation of the equation
${\cal O} |f\rangle = |d\rangle$ is given by

\begin{equation}
\sum_{j\beta} {\cal O}_{i\alpha,j\beta} f_{j\beta} = d_{i\alpha}\,.
\end{equation}
Making use of Eqs.~(\ref{redf}) and (\ref{inverse}) we find
\begin{equation}
\sum_j {\cal O}_{ij}^{(\lambda)} f_j^{(\lambda)} = d_i^{(\lambda)}~,
\label{redmat}
\end{equation}
where
\begin{eqnarray}
d_i^{(\lambda)} &=& \sum_{\alpha} d_{i\alpha} 
v_{\alpha}^{(\lambda)\star} \label{dlambda}
\\ {\cal O}_{ij}^{(\lambda)} &=& 
\sum_{\alpha \beta} v_{\alpha}^{(\lambda)\star} 
{\cal O}_{i\alpha,j\beta} v_{\beta}^{(\lambda)}~.
\end{eqnarray}
In obtaining Eq.~(\ref{redmat}) we have used the fact that the operator
${\cal O}$ is diagonal in the $\Gamma(C_4)$ basis.
It is clear from Eq.~(\ref{redmat}) that each of the eigenvalues, 
$\lambda$, defines an independent matrix problem.
This is a direct consequence of the fact that we are working in the 
symmetrized basis of the representation $\Gamma$ of $C_4$.
Furthermore, suppose ${\cal O} = {\cal PQ}$ with ${\cal P}$ and 
${\cal Q}$ both invariant under $C_4$.  It is easy to show that

\begin{equation}
{\cal O}_{ij}^{(\lambda)} = \sum_k {\cal P}_{ik}^{(\lambda)} 
{\cal Q}_{kj}^{(\lambda)}~,
\end{equation}
which means that for a given $\lambda$, we need deal only with the
shell matrices rather than the matrices defined over all
four quadrants of the reciprocal space.

Let us now apply what we have learned to Eq.~(\ref{inhomoeigenvalue})
in the ${\bf q} = 0$ limit. It is readily verified that
$A_{{\bf G}{\bf G}'}$, $B_{{\bf G}{\bf G}'}$, and 
${\tilde M}_{{\bf G}{\bf G}'}$, are invariant under the symmetry
group $C_4$.  Therefore, Eq.~(\ref{inhomoeigenvalue}) can be
immediately cast in terms of the symmetrized functions, {\em viz.,}

\begin{equation}
\omega (\tilde{\omega}^2 - \omega_{c}^{2}) c_{i}^{(\lambda)} -
\tilde{\omega} \sum_k B^{(\lambda)}_{ij}\tilde{M}^{(\lambda)}_{jk} 
c_{k}^{(\lambda)} -
\omega_c \sum_k A^{(\lambda)}_{ij}\tilde{M}^{(\lambda)}_{jk}
c_{k}^{(\lambda)} = b_i^{(\lambda)}~, 
\label{redinhomoeigenvalue}
\end{equation}
where we recall that summations are over the shell index.
In obtaining (\ref{redinhomoeigenvalue}), we have used the relation

\begin{equation}
f_i^{(\lambda)} = \sum_j \tilde{M}_{ij}^{(\lambda)} c_j^{(\lambda)}\,.
\end{equation}
In principle, Eq.~(\ref{redinhomoeigenvalue}) must be solved for each 
of the four eigenvalues of $C_4$.  However, as we mentioned above,
if we are only interested in the FIR-active modes of the system, we can
project into the invariant subspace of $\Gamma_5$, and only have to 
consider at most, $\lambda^{(3)}$ and $\lambda^{(4)}$.
This can be made more transparent as follows.
The reciprocal lattice vector ${\bf G}_{i\alpha}$ is given by
(see Fig. \ref{appfig1})

\begin{equation}
{\bf G}_{i\alpha} = |G_{i1}|\left ( {\rm cos}(\phi_{i\alpha}),
{\rm sin}(\phi_{i\alpha})\right )~,
\label{gia}
\end{equation}
where

\begin{equation}
\phi_{i\alpha} = \theta_i + \frac{(\alpha -1)\pi}{2}~,
\end{equation}
and
\begin{equation}
\theta_i \equiv {\rm cos}^{-1} \left( \frac{G_{i1,x}}{|G_{i1}|} 
\right )~.
\end{equation}
The inhomogeneous vector on the right-hand-side of 
Eq.~(\ref{redinhomoeigenvalue}) is determined from Eq.~(\ref{dlambda}).
Using Eq.~(\ref{gia}) in Eq.~(\ref{inhomovec}), we obtain the 
expression
$b_{i\alpha}$ 

\begin{equation}
b_{i\alpha} = -E_0 \left (i\tilde{\omega}  {\rm cos}(\phi_{i\alpha}) - 
\omega_c {\rm sin}(\phi_{i\alpha})\right )|G_{i1}|\psi_0[{\bf G}_{i1}]~,
\end{equation}
which yields

\begin{eqnarray}
&b_i^{(\lambda=1)}& = 0 \nonumber \\
&b_i^{(\lambda=-1)}&=  \nonumber 0\\
&b_i^{(\lambda=i)}& = -i|G_{i1}|\psi_0[{\bf G}_{i1}]E_0 \left (
\tilde{\omega} + \omega_c\right ) e^{i\theta_i}  \nonumber \\
&b_i^{(\lambda=-i)}& = -i|G_{i1}|\psi_0[{\bf G}_{i1}]E_0 \left (
\tilde{\omega} - \omega_c\right ) e^{-i\theta_i} = 
b_i^{(\lambda=i)\star}(-\omega)~.
\label{bilambda}
\end{eqnarray}
Therefore, as expected, we only require the $\lambda = \pm i$ 
eigenvalues to fully describe the FIR response.  In particular, the
determination of the power absorption is reduced to a solution of 
Eq.~(\ref{redinhomoeigenvalue}) for $\lambda = \pm i$. Using these
solutions in Eq.~(\ref{j_G=0}),
we finally obtain the following explicit
expression of the power absorption:
\begin{eqnarray}
\frac{\langle P \rangle_t}{A} &=& \frac{1}{2}E_0 {\rm Re}
\left [ \frac{i\tilde{\omega}n_0[{\bf G}=0] E_0}{\tilde{\omega}^2 
- \omega_c^2} - \frac{4}{\tilde{\omega} - \omega_c}
\sum_i |G_{i1}| \psi_0[{\bf G}_{i1}] \sum_j \tilde{M}_{ij}^{(\lambda=i)}
c_j^{(\lambda=i)} e^{-i\theta_i} \right . \nonumber \\
&-& \left . \frac{4}{\tilde{\omega} + \omega_c}
\sum_i |G_{i1}| \psi_0[{\bf G}_{i1}] \sum_j 
\tilde{M}_{ij}^{(\lambda=-i)} c_j^{(\lambda=-i)} e^{i\theta_i}
\right ]~.
\end{eqnarray}

To obtain the normal mode frequencies, we use the homogeneous version of
Eq.~(\ref{redinhomoeigenvalue}) with $\gamma =0$. In this case, all {\it
positive} eigenvalues for $\lambda = \pm i$ can be obtained from the
full set of eigenvalues for $\lambda = i$. The negative eigenvalues
simply correspond to the positive eigenvalues for $\lambda = -i$.
This is the method used to generate the magnetic dispersion of the
FIR-active modes shown in Figs.~2, 4 and 5.

\begin{figure}
%\begin{center}
%\leavevmode
%\hbox{%
%\epsfxsize=8cm
%\epsffile{paradot.ps}}
%\end{center}
\caption{Equilibrium density distributions for various confinement
geometries: (a)
parabolic confinement, (b) radial anharmonicity and (c) noncircular
symmetry. Parameters defining the confinement potential are given in 
the text. In all cases the dots are arranged on a square lattice with
$a_x = a_y = 1000$ nm, and contain $N=210$ electrons.}
\label{density}
\end{figure}

\begin{figure}
%\begin{center}
%\leavevmode
%\hbox{%
%\epsfxsize=14cm
%\epsffile{B-dispersion-paradot.ps}}
%\end{center}
\caption{The magnetic dispersion for an array of well separated 
parabolically confined dots.
The solid circles are the numerical solutions to Eq.~(\ref{eigenvalue}) 
and the solid curves are fits to Eq.~(\ref{wpm}).
The figure inset shows the calculated FIR power absorption for a range
of magnetic fields. The parameters defining the dots are the same
as used to generate Fig.~\ref{density}(a).}
\label{B-paradot}
\end{figure}

\begin{figure}
%\begin{center}
%\leavevmode
%\hbox{%
%\epsfxsize=14cm
%\epsffile{B-dispersion-closeparadot.ps}}
%\end{center}
\caption{The magnetic dispersion for an array of closely spaced dots.
The only difference between this figure and Fig.~\ref{B-paradot} is
a reduction of the lattice periodicity: $a_x = a_y = 600$ nm.
The consequences of this shorter periodicity are a softening of the
$\omega_0$ frequency and the appearance of small anticrossings.
Inset:  A expanded view of the boxed region 
$\approx (1.4{\rm T},39~{\rm cm}^{-1})$ clearly showing the
mode anticrossings.}
\label{B-closeparadot}
\end{figure}

%\begin{figure}
%\begin{center}
%\leavevmode
%\hbox{%
%\epsfxsize=8cm
%\epsffile{r4dot.ps}}
%\end{center}
%\caption{The equilibrium density distribution
%for an array of anharmonic dots. 
%The parameters are chosen so as to fit the experimental zero field frequency
%in Ref.[9]: $\eta=1$, 
%$\omega_0 = 20~{\rm cm}^{-1}$, $N=210$ and $a_x = a_y = 1000$ nm.}
%\label{density-r4dot}
%\end{figure}

\begin{figure}
%\begin{center}
%\leavevmode
%\hbox{%
%\epsfxsize=14cm
%\epsffile{B-dispersion-r4dot.ps}}
%\end{center}
\caption{The magnetic dispersion for an array of radially anharmonic 
dots. The solid circles are the numerical solutions to 
Eq.~(\ref{eigenvalue}) and
the solid curves are fits to  Eq.~(\ref{wpm}).
The parameters are the same as those used to generate
Fig.~\ref{density}(b). The inset shows the calculated FIR power 
absorption.}
\label{B-r4dot}
\end{figure}

%\begin{figure}
%\begin{center}
%\leavevmode
%\hbox{%
%\epsfxsize=12cm
%\epsffile{power_r4dot.ps}}
%\end{center}
%\caption{The calculated FIR power absorption for an array of anharmonic
%dots.  The same parameters as in Fig.~\ref{density}(b) are
%used to generate this figure.}
%\label{pwrr4dot}
%\end{figure}

%\begin{figure}
%\begin{center}
%\leavevmode
%\hbox{%
%\epsfxsize=8cm
%\epsffile{temp.ps}}
%\end{center}
%\caption{The equilibrium density distribution
%for an array of noncircular dots. 
%The parameters are chosen so as to fit the experimental zero field frequency
%in Ref. [9]: $\eta=1$, 
%$\omega_0 = 20~{\rm cm}^{-1}$, $\varepsilon=0.4$, 
%$N=210$ and $a_x = a_y = 1000$ nm.}
%\label{density-noncirdot}
%\end{figure}

\begin{figure}
%\begin{center}
%\leavevmode
%\hbox{%
%\epsfxsize=14cm
%\epsffile{B-dispersion-noncirdot.ps}}
%\end{center}
\caption{The magnetic dispersion for an array of noncircular dots. The 
solid circles are the numerical solutions to Eq.(\ref{eigenvalue}) and
the solid curves are fits to  Eq.~(\ref{wpm}). The same parameters as 
in Fig.~\ref{density}(c) are used to generate this figure.
The inset shows the calculated FIR power absorption.
The size and position of the pronounced anticrossing at
$\omega \approx 40~{\rm cm}^{1}$ ($\uparrow$) are in quantitative
agreement with experiment. 
For $B > 3 T$, the power absorption is almost indistinguishable from
Fig.~\ref{B-r4dot}.}
\label{B-noncirc}
\end{figure}

\begin{figure}
%\begin{center}
%\leavevmode
%\hbox{%
%\epsfxsize=12cm
%\epsffile{reciprocal-lattice.ps}}
%\end{center}
\caption{A schematic representation of the truncated 2D reciprocal 
lattice with point group symmetry $C_{4v}$. The arrows
are an example of a shell of vectors $\{|{\bf G}_\alpha \rangle \}$ 
which are generated from ${\bf G}_1$ by the symmetry operation $C_4$.
The solid circles illustrate the set of ${\bf G}$ vectors 
used in the symmetry-reduced calculation; a square array of points is
retained to facilitate the use of fast Fourier transforms.}
\label{appfig1}
\end{figure}

\begin{table}
\begin{center}
\begin{tabular}{ccccrrrrrrrrc}

&$C_{4v}$& &$E$& &$2C_4$& &$C_4^2$& &$2m$& &$2\sigma$&\\ \hline
&$\Gamma_1$& &1& &1& &1& &1& &1&\\
&$\Gamma_2$& &1& &$-1$& &1& &1& &$-1$&\\
&$\Gamma_3$& &1& &1& &1& &$-1$& &$-1$& \\
&$\Gamma_4$& &1& &$-1$& &1& &$-1$& &1&\\
&$\Gamma_5$& &2& &0& &$-1$& &0& &0& \\
\end{tabular}
\end{center}
\caption{Character table for the irreducible representations of the 
point group $C_{4v}$.}
\label{charc4v}
\end{table}

\begin{table}
\begin{center}
\begin{tabular}{ccccrrrrrrc}

&$C_4$& &$E$& &$C_4$& &$C_4^2$& &$C_4^3$&\\ \hline
&$D_1$& &1& &1& &1& &1&\\
&$D_2$& &1& &$-1$& &1& &$-1$& \\
&$D_3$& &1& &$i$& &$-1$& &$-i$& \\
&$D_4$& &1& &$-i$& &$-1$& &$i$& \\
\end{tabular}
\end{center}
\caption{Character table for the irreducible representations of the 
point group $C_4$.}
\label{charc4}
\end{table}

\begin{table}
\begin{center}
\begin{tabular}{ccccrrrrrrc}

&$m^\pi$& &$E$& &$C_4$& &$C_4^2$& &$C_4^3$&\\ \hline
&$0^+$& &1& &1& &1& &1&\\
&$0^-$& &1& &1& &1& &1&\\
&$1^+$& &1& &$i$& &$-1$& &$-i$& \\
&$1^-$& &1& &$-i$& &$-1$& &$i$& \\
&$2^+$& &1& &$-1$& &1& &$-1$& \\
&$2^-$& &1& &$-1$& &1& &$-1$& \\
&$3^+$& &1& &$-i$& &$-1$& &$i$& \\
&$3^-$& &1& &$i$& &$-1$& &$-i$&

\end{tabular}
\end{center}
\caption{Compound characters for some of the $SO(2)$ representations of
the point group $C_4$.}
\label{charso2}
\end{table}

\end{document}